\documentclass[authorversion,screen]{acmart}

\AtBeginDocument{%
  }

 \acmDOI{}

\setcopyright{rightsretained}

\acmConference[CHI'25]{ACM CHI 2025 Workshop WS 12: Scaling Distributed Collaboration in Mixed Reality}{April 26,
  2025}{Yokohama, Japan}

\begin{document}


\title{Managing Information Overload in Large-Scale Distributed Mixed-Reality Meetings}

\author{Katja Krug}
\email{katjakrug@acm.org}
\orcid{0000-0003-4800-6287}
\authornote{Also with Centre for Tactile Internet with Human-in-the-Loop (CeTI)}

\author{Wolfgang Büschel}
\email{bueschel@acm.org}
\orcid{0000-0002-3548-723X}

\author{Mats Ole Ellenberg}
\email{mellenberg@acm.org}
\orcid{0000-0002-0354-1537}
\authornotemark[1]

\affiliation{%
  \institution{ \\ Interactive Media Lab Dresden, TUD Dresden University of Technology}
  \city{Dresden}
  \country{Germany}
}
\renewcommand{\shortauthors}{Krug et al.}

\begin{abstract}
 Large-scale distributed mixed-reality meetings involve many people and their audiovisual representations. These collaborative environments can introduce challenges such as sensory overload, cognitive strain, and social fatigue. In this paper, we discuss how the unique adaptability of Mixed Reality can be leveraged to weaken these stressors by managing information overload. 
\end{abstract}


\received{14 March 2025} 
\received[accepted]{24 March 2025}

\maketitle

\section{Introduction and Background}
In recent years, commercially available head-mounted displays (HMDs) have brought Virtual and Mixed Reality (VR and MR) technology into homes and workplaces.
With video see-through MR being supported by most current-gen devices, applications can blend physical and virtual worlds, paving the way for immersive, mixed-reality collaborations, where distributed users come together as virtual avatars.  
In a future where such devices are more widespread, supporting usage scenarios with larger groups of users will become crucial.
One fundamental challenge of scaling up these mixed-reality collaborations is to support human cognition and limit information overload.
Real-world, large-scale meetings with dozens or even hundreds of participants are often seen as stressful and unpleasant because of factors such as environmental overstimulation, i.e., information overload and noise, which can lead to disorientation and simplified information processing \cite{suedfeld1985stressful}, hindering complex collaboration.
The social overload of large-scale meetings can further induce fatigue and reduced engagement \cite{zhang2020investigating}.
In MR, perceptual issues such as mismatching virtual and real spaces \cite{drascic1996perceptual}, and insufficient HMD ergonomics \cite{chen2021human} can be further additional stressors.


In the well-established field of research of adaptive user interfaces, many efforts have been made to support cognition and optimize interaction, e.g., data in MR interfaces can be filtered or clustered to reduce information density \cite{tatzgern2016adaptive} and semantic level-of-detail rendering \cite{wysopal2023level} can be used to optimize content to the available screen space.
Context-aware adaptations for Mixed Reality have been investigated, among others, by Lindlbauer et al.~\cite{lindlbauer2019context} and Lie et al.~\cite{li2024situationadapt}.
While most of these adaptations at least implicitly support cognition, some systems directly aim at supporting users in challenging contexts, such as when driving \cite{wen2024adaptivevoice}.
However, we see untapped potential in researching adaptations for large-scale distributed MR systems specifically, where any adaptive change to how and what content is shown can not be seen in isolation. Instead, the impact on collaboration and communication has to be considered as well.

In this paper, we explore and discuss such context- and preference-dependent adaptations and techniques in relation to the presentation and distribution of individuals in large-scale MR meetings. We further describe methods to control such functionality, including user-driven and automatic approaches.

\section{Adaptations and Techniques}
In the following, we list examples of techniques MR systems could employ to reduce cognitive overload and support collaboration in large-scale MR meetings. 
This list is not comprehensive, instead it is aimed to be an inspirational starting point for discussions about MR adaptations. 
Many of these techniques are based on the perceived relevance of people to individual users. 
This relevance is highly personal and can mean a pre-determined person of interest, an active speaker, a person in one's line of sight, or a potential collaborator, etc. 
We further discuss these dependencies in \autoref{methods}.

\subsection{Visual Rendering, Filtering, and Grouping of Avatars}
\label{adaptations:visual}
Enhancing focus and reducing distractions, adaptive visual filtering techniques can be employed to prioritize information and minimize visual complexity in distributed MR meetings. 
One approach is \textbf{idle avatar reduction}, where inactive avatars could fade out or be replaced by minimal icons until they re-engage.
To reduce the visual noise of excessive movement, \textbf{key gesture filtering} could limit animations to important communicative gestures like pointing or nodding, and otherwise only display subtle idle animations to avoid the eeriness of completely still avatars. 
\textbf{Relevance-based avatar scaling}, i.e., adjusting avatar size based on importance, could make key individuals more prominent and easy to find while reducing visual clutter from bystanders.
Similarly, \textbf{selective rendering and transparency} can de-emphasize less relevant avatars by abstracting or fading them out, and enhance important ones by highlighting them with a certain color or additional visualizations, or adding fine details like facial expressions.
\textbf{Visual avatar grouping} can help structuring the social space by clustering related individuals together, e.g., people from the same work group. 
This can be done through proximity or adding shared visual traits such as color-coding or uniform clothing.
Clusters of avatars could also be combined into cohesive "crowd blobs" that only disperse upon focus, or into a single representative avatar that splits into multiple when triggered. 

\subsection{Audio Management}
To avoid users becoming overwhelmed by the competing voices and background noise of large crowds, adaptive audio management techniques can help prioritizing important sounds and suppress distractions.
\textbf{Audio prioritization and speech isolation} could emphasize relevant voices while fading out or muting less important ones. 
To prevent auditory overload, \textbf{conversational blur} can merge excessive background voices into a subtle, incomprehensible background noise, ensuring that users remain aware of if and where active conversations are taking place without being overwhelmed. 
The system can then dynamically clarify speech, e.g., when only few people are talking and most are listening, or when important speakers are talking.
Additionally, \textbf{adaptive noise suppression} can help regulating background sound and maintain group awareness, e.g., as a large group gradually quiets down, the system can dynamically strengthen the noise suppression to facilitate upcoming announcements. 
    
    
\subsection{Environment Partitioning and Crowd Distribution}
Unlike physical environments, Mixed Reality spaces are flexible and can dynamically adapt to user needs, improving coordination and cognitive ease.  
\textbf{Dynamic room partitioning} can help distribute avatars evenly to prevent overcrowding by adjusting the layout of the virtual space.
\textbf{Virtual borders} are an alternative to the visual grouping mentioned in \autoref{adaptations:visual}, emphasizing these partitions by erecting virtual walls or subtle floor markings that encircle cohesive groups, which could trigger functionality upon crossing, like making certain conversations audible. 
In-between, \textbf{generated pathways} can ensure smooth navigation by establishing routes where movement is allowed and gathering is restricted.
To further facilitate discussions, \textbf{auto grouping} could detect common topics and dynamically form discussion clusters, gently relocating them to designated areas.  
Additionally, \textbf{elastic room scaling} can leverage MR's lack of physical constraints by expanding the virtual space past the physical one, subtly pushing less relevant groups outwards to maintain focus con key conversations.
\textbf{Density balancing encouragements}, in the form of subtle visual cues,  could encourage users to spread out in crowded areas, promoting a more comfortable and breathable virtual space.
\textbf{Dedicated privacy zones} could allow users to retreat into a quiet, shielded area for distraction-free conversation. 
For those feeling overwhelmed, a \textbf{virtual palate cleanser environment} can offer a calming space to mentally reset, similar to stepping outside for fresh air. 
In extreme cases of mental overload, a \textbf{quick escape mechanism} should allow users to temporarily leave the virtual environment without exiting the application, keeping their avatar in an idle state, labeled for return.

\subsection{AI Assisted Attention Management}
In real-life interactions, body language, context, and physical proximity can naturally guide attention. 
In virtual environments, these cues are limited, requiring additional guidance support to avoid cognitive overload.  
The system could provide on-demand \textbf{conversation summaries}, i.e., brief text or audio recaps of relevant conversations, helping users decide if they want to join a discussion group. 
Similarly, \textbf{content labels} could continuously display dynamically adapting key conversation topics above groups, along with optional icons indicating the emotional tone of the conversation, e,g,, friendly, focused, or intense.
To aid the discovery of relevant discussions, suggestions for relevant groups or individuals can be made through \textbf{engagement predictions}, generated based on user interests, profiles, or prior interactions.  
Additionally, \textbf{keyword-based filtering} can refine the ongoing auditory input, ensuring that users only hear conversations containing relevant terms that are either user-defined or AI generated. 
\textbf{Smart notifications} can dynamically alert and potentially guide users to new participants or relevant discussions, timing notifications strategically to minimize disruptions while enhancing awareness.


\section{Adaptation Method}
\label{methods}
The effectiveness of adaptive techniques for managing information overload in MR depends not only on what is adapted but also on how these adaptations are triggered.
These features range from fully user-driven (full control) to fully system-driven (automation without user intervention).  

\subsection{User-Driven}
Here, users manually trigger and customize adaptions. This provides maximum control and transparency but requires a conscious effort and decision-making. User-driven systems can become complex and hard to generalize.  

\textbf{\textit{Explicit User Decisions:}}
The user maintains full control throughout and actively makes decisions, such as determining people of interest and their rendering adaptations, or hand-drawing privacy bubbles.  
While the customizability makes the system highly inclusive and accessible, this approach risks decision fatigue \cite{baumeister2018ego} and cognitive overload, since the system quickly becomes complex and hard to operate. 

\textbf{\textit{Implicit System Reactions:}}
The adaptations respond automatically to user actions, such as adjusting avatar relevance based on the user's current Field of View (FoV) or initializing calm-down techniques when detecting a rising heart rate.
This reduces manual user effort while maintaining some level of control. However, compared to the explicit decisions, this raises privacy and security concerns, as it requires the processing of sensitive data. It also risks misinterpretation, such as unintentional audio shifts due to FoV movement (Midas touch effect). 

\subsection{System-Driven}
In system-driven adaptations, the system autonomously adjusts the settings in real time based on what it learns from user behavior, social dynamics or environmental factors. 
This allows for seamless integration and reduced cognitive load, but might cause a feeling of loss of control and disorientation, raises privacy concerns, and might even negatively influence collaboration, if the system's predictions and conclusions are inaccurate.

\textbf{\textit{User Parameter Dependent}}
The system automatically personalizes adaptions based on the user profile, such as highlighting co-workers, emphasizing discussion contributions by higher ranking roles, alerting users about conversations adjacent to their expertise. 
This facilitates networking while reducing its awkwardness, but risks echo chambers and neglects potentially shifting goals and interests during conversations.  

\textbf{\textit{Context-Triggered}}
Adaptations respond to the user's current context, such as visually emphasizing a speaker with relevant contributions to the topic, or muting background noise when the user is part of a discussion group.
This subtly improves focus and enables smooth transitions between contexts, but may feel unexpected if the user is not aware of the context changing. 
The system might also misinterpreted the context.

\textbf{\textit{Socially-Adaptive}}
The adaptations are based on interpersonal relationships and group behavior, such as visually emphasizing a speaker that is being looked at by many people, or clustering habitual conversation partners and guiding them towards each other in the crowd.
This helps strengthen interpersonal relationships and leadership emergence, but can reinforce bias, suppress less prominent voices and limit new connections. 

\section{Conclusion}
Large-scale MR meetings can induce diverse cognitive challenges from information overload to social fatigue. 
To address these, we explored adaptations and techniques that dynamically manage sensory input and support communication. 
These techniques exist along the user agency spectrum, ranging from complete user control to fully automated adaptations.
While the user-driven methods offer transparency and customization, they can lead to decision fatigue and interaction complexity. 
On the other hand, fully automated solutions reduce cognitive effort but risk misinterpretation, loss of control, and raise privacy concerns. 
Hybrid user-system solutions could act as a promising middle-ground between these ends of the spectrum, balancing automation with user agency. 
AI could suggest optimizations that users can refine or override. 
Users can make active selections of people and spaces and AI adaptations honor and emphasize these selections. 
Adaptations should be unobtrusive, assistive, and customizable. They should enhance collaboration and keep users empowered without dictating their experience. 
Transparency is crucial and users should always understand how and why their environment is adapting. 
The thoughtful design of these MR adaptations can reduce cognitive strain and create more comfortable and intuitive distributed large-scale MR meetings. 

\begin{acks}
This work has been funded by the German Research Foundation (DFG, Deutsche For-schungsgemeinschaft) as part of Germany’s Excellence Strategy – EXC 2050/1 – Project ID 390696704 – Cluster of Excellence “Centre for Tactile Internet with Human-in-the-Loop” (CeTI) of Technische Universität Dresden. The authors further acknowledge the financial support by the Federal Ministry of Education and Research of Germany in the programme of “Souverän. Digital. Vernetzt.”. Joint project 6G-life, project identification number: 16KISK001K.
\end{acks}



\bibliographystyle{ACM-Reference-Format}
\bibliography{references}

\end{document}